\documentclass[twocolumn,showpacs,preprintnumbers,amsmath,amssymb]{revtex4}

\usepackage{graphicx}
\usepackage{dcolumn}
\usepackage{bm}

\begin{document}


\title{No Maxwell Electromagnetic Wave Field Excited In Cloaked Concealment}

\author{Ganquan Xie}
 \altaffiliation[Also at ]{GL Geophysical Laboratory, USA, glhua@glgeo.com}
\author{Jianhua Li, Lee Xie, Feng Xie}%
 \email{GLGANQUAN@GLGEO.COM}
\affiliation{%
GL Geophysical Laboratory, USA
}%

\hfill\break

\date{\today}

\begin{abstract}
The GL electromagnetic (EM) modeling is used to simulate the 3D EM
full wave field propagation through cloaks. The 3D GL simulation of the EM 
wave field excited by a point source outside of the cloaks has been 
done. The simulation of the EM wave field from the point source excitation 
inside of the cloak device is presented in this paper. 
By using the GL modeling simulation, 
we found a phenomenon that 
there is no Maxwell EM wave field which is excited by nonzero local sources inside 
of  the single layer cloaked concealment.  The theoretical proof of the phenomenon 
by GL method is proposed in this paper. The GL method is fully 
different from the conventional methods. The GL method has 
double abilities of  the theoretical analysis and numerical simulations to research the 
physical process and cloak metamaterial properties that is exhibited in this paper.
\end{abstract}

\maketitle

\section{\label{sec:level1}INTRODUCTION} 
 For the point source located outside of the cloak, 
the 3D Global and Local field (GL) EM modeling method [1-3]
 has been used to simulate the 3D full EM wave field propagation 
through the cloaks. Our simulations and theoretical analysis in [1] 
verify the ideal cloak functions [4]. The EM wave field propagation outside 
of cloak does not penetrate into the concealment and never be
disturbed by the cloaks. There  are several papers to simulate 
the plane wave propagation through the cloak from outside 
of the cloak [5-7]. The plane wave is excited by plane source 
which can not be located inside of the cloak or concealment. 
The cloak  simulation of the point sources and the local sources inside of 
the cloak is absent. 
in the mose published papers. The GL modeling simulations of the EM 
wave field through cloak and excited by the nonzero local sources 
inside of the cloak is presented in this paper. Moreover, 
by GL EM method simulation of the nonzero local sources 
inside of the concealment, we found a phenomenon that 
there exists no Maxwell EM wave field excited by nonzero local sources 
inside of the single layer cloaked concealment. The phenomenon is 
proved by the GL method theoretical analysis and the EM integral equation [1-3]. 
In paper [9] and [10], authors studied the effect on invisibility of active devices 
inside the cloaked region.
Our statement is that "there exists no Maxwell EM wave field
can be excited by nonzero local sources inside of the single layer cloaked concealment
with normal materials".
The detailed proof of the statement and 3D GL simulations are presented in this paper.

Our GL method is fully different from conventional methods for cloak 
and physical and science simulations. It has advantages over the
conventional methods. The GL method consistent combines the theoretical 
analytical and numerical method together.
In the GL modeling, there is no big matrix equation to solve 
and no absorption condition on artificial boundary to truncate 
infinite domain. The method is a significant physical scattering 
process. The finite inhomogeneous domain is divided into a set 
of small sub domains. The interaction between the global field 
and anomalous material polarization field in the sub domain 
causes a local scattering wave field. The local scattering wave 
field updates the global wave field by an integral equation. 
Once all sub domains are scattered, the wave field in the 
inhomogeneous anomalous materials will be obtained. 
Therefore, the GL method can be used to
both of theoretical analysis and numerical simulation for physical and chemical
phenomena and process.

The description arrangement of the paper is as follows. 
The introduction is described in Section 1. The EM 
integral equation is proposed in Section 2. The Global 
and Local EM field modeling is proposed in Section 3. 
The phenomenon that there is no the Maxwell EM 
wave field excited by the local sources inside of the concealment is proved in Section 4. 
The GL modeling simulations of the EM wave field propagation excited by the local source 
inside of  the cloak are presented in Section 5. We conclude this paper in Section 6.

\section{\label{sec:level1}3D ELECTROMAGNETIC INTEGRAL EQUATION}

We have proposed the 3D EM integral equation in frequency domain 
in papers [1] and [2].  In this section, we propose the EM integral equation in time domain as follows:

\begin{equation}
\begin{array}{l}
 \left[ {\begin{array}{*{20}c}
   {E(r,t)}  \\
   {H(r,t)}  \\
\end{array}} \right] = \left[ {\begin{array}{*{20}c}
   {E_b (r,t)}  \\
   {H_b (r,t)}  \\
\end{array}} \right] \\ 
  + \int\limits_\Omega  {G_{E,H}^{J,M} (r',r,t) * _t \delta \left[ {D(r')} \right]\left[ {\begin{array}{*{20}c}
   {E_b (r',t)}  \\
   {H_b (r',t)}  \\
\end{array}} \right]dr'}, \\ 
 \end{array}
\end{equation}
where
\begin{equation}
G_{E,H}^{J,M} \left( {r',r,t} \right) = \left[ {\begin{array}{*{20}c}
   {E^J \left( {r',r,t} \right)} & {H^J \left( {r',r,t} \right)}  \\
   {E^M \left( {r',r,t} \right)} & {H^M \left( {r',r,t} \right)}  \\
\end{array}} \right],
\end{equation}
${E(r,t)}$ is the electric field, ${H(r,t)}$ is the magnetic field, $E_b (r,t)$ and
 $H_b (r,t)$ is the incident electric and magnetic field
 in the background medium, $E^J (r',r,t)$ is electric Green's tensor, $H^J (r',r,t)$ is 
 magnetic Green's tensor, they are excited by the point impulse current source,
$E^M(r',r,t)$ and $H^M (r',r,t)$ are electric and magnetic Green's tensor, respectively,
 they are excited by the point impulse magnetic moment source,
 $ * _t$ is convolution with respect to t,
$\delta \left[ D \right]$ is the electromagnetic material  parameter variation matrix,
\begin{equation}
\begin{array}{l}
 \delta \left[ D \right] = \left[ {\begin{array}{*{20}c}
   {\delta D_{11} } & 0  \\
   0 & {\delta D_{22} }  \\
\end{array}} \right], \\ 
 \delta D_{11}  = (\bar \sigma (r) - \sigma _b I) + (\bar \varepsilon (r) - \varepsilon _b I)\frac{\partial }{{\partial t}}, \\ 
 \delta D_{22}  = (\bar \mu (r) - \mu _b I)\frac{\partial }{{\partial t}}, \\ 
 \end{array}
\end{equation}
$\delta D_{11} $ and $\delta D_{22}  $ are a $3\times 3$ symmetry, 
inhomogeneous diagonal matrix  for  the isotropic material,
for anisotropic material, they are an inhomogeneous diagonal or full matrix, 
$I$ is a $3\times 3$ unit matrix, $\bar \sigma (r)$ is the conductivity tensor,
$\bar \varepsilon (r)$ is the dielectric tensor, $\bar \mu (r)$ is susceptibility tensor which can be dispersive
parameters depend on the angular frequency $\omega$, 
$\sigma _b$ is the conductivity,  $\varepsilon_b $
is the permittivity, $ \mu _b$ is the permeability in the background free space, $\Omega$
is the finite domain in which the parameter variation matrix $\delta \left[ D \right] \ne 0,$
the $(\bar \varepsilon (r) - \varepsilon _b I)E $ 
is the electric polarization, and $(\bar \mu (r) - \mu _b I)H $ is the magnetization.

\section{\label{sec:level1}3D GL EM MODELING}
We propose the GL EM modeling based on the EM integral equation (1) in the time space
domain.

 (3.1)	The domain $\Omega$ is divided into a set of $N$ sub domains,$\{\Omega_k\}$, 
such that $\Omega  = \bigcup\limits_{k = 1}^N {\Omega _k }$. The division can be mesh or meshless.

(3.2)  When $k=0$, let
$E_0 (r,t)$ and $H_0 (r,t)$ are the analytical global field, $E_0^J (r',r,t)$, $H_0^J (r',r,t)$, $E_0^M (r',r,t)$, and
$H_0^M (r',r,t)$ are the analytical global Green's tensor in the background medium. By induction, suppose that
$E_{k-1} (r,t)$, $H_{k-1} (r,t)$, $E_{k-1}^J (r',r,t)$, $H_{k-1}^J (r',r,t)$, $E_{k-1}^M (r',r,t)$, and
$H_{k-1}^M (r',r,t)$ are calculated in the $(k-1)^{th}$ step in the subdomain $\Omega_{k-1}$.

(3.3) In $\{\Omega_k\}$, upon substituting $E_{k-1} (r,t)$, $H_{k-1} (r,t)$, $E_{k-1}^J (r',r,t)$, $H_{k-1}^J (r',r,t)$, $E_{k-1}^M (r',r,t)$, and
$H_{k-1}^M (r',r,t)$ into the integral equation (1),  the EM Green's tensor integral equation (1)
in $\Omega_{k}$ is reduced into $6\times 6$ matrix equations. By solving the $6\times 6$  matrix equations, 
we obtain the Green's tensor field  $E_{k}^J (r',r,t)$, $H_{k}^J (r',r,t)$, $E_{k}^M (r',r,t)$, and
$H_{k}^M (r',r,t)$.

(3.4) According to the  integral equation (1), the electromagnetic field $E_{k} (r,t)$ and $H_{k} (r,t)$ are updated by 
the interaction scattering field between the Green's tensor and local polarization and magnetization in the subdomain $\Omega_{k}$ as follows,
\begin{equation}
\begin{array}{l}
 \left[ {\begin{array}{*{20}c}
   {E_k (r,t)}  \\
   {H_k (r,t)}  \\
\end{array}} \right] = \left[ {\begin{array}{*{20}c}
   {E_{k - 1} (r,t)}  \\
   {H_{k - 1} (r,t)}  \\
\end{array}} \right] \\ 
  + \int\limits_{\Omega_k}  {\left\{ {\left[ {\begin{array}{*{20}c}
   {E_k^J (r',r,t)} & {H_k^J (r',r,t)}  \\
   {E_k^M (r',r,t)} & {H_k^M (r',r,t)}  \\
\end{array}} \right]} \right.}  \\ 
 \left. { * _t \delta \left[ {D(r')} \right]\left[ {\begin{array}{*{20}c}
   {E_{k - 1} (r',t)}  \\
   {H_{k - 1} (r',t)}  \\
\end{array}} \right]} \right\}dr' \\ 
 \end{array}
\end{equation}

(3.5) The steps (3.2) and (3.4) form a finite iteration, $k = 1,2, \cdots, N$,
the $E_N \left( r,t \right)$ and $H_N \left( r,t \right)$
are the electromagnetic field  of the GL modeling method. The GL
electromagnetic field modeling in the time space domain is short named as
GLT method.

The GL EM modeling in the space frequency domain is proposed in the paper \cite{2}, 
we call the GL modeling in frequency domain as GLF method. 

\section{\label{sec:level1}NO MAXWELL ELECTROMAGNETIC  WAVE  FIELD
EXCITED IN CLOAKED CONCEALMENT}

$\textbf{Theorem:}$ Suppose that a 3D anisotropic inhomogeneous closed 
strip cloak domain separates the whole 3D space into three sub domains, 
one is the cloak domain $\Omega _c$ with the cloak material; the second one is the cloaked 
concealment domain $\Omega _d$ with normal EM materials; other one is the 
free space outside of the cloak. If the Maxwell EM wave field excited 
by a point source outside of the concealment to be vanished
in inside of the concealment, then there is no Maxwell EM wave field excited 
by the local sources inside of the cloaked concealment.

The Maxwell EM wave field is the  EM wave field which satisfies the Maxwell equation
and continuous interface boundary conditions. We call the Maxwell EM wave field 
 as the EM wave  field and use inverse process to prove the theorem as follows:
Suppose that there exists Maxwell EM wave field excited by the local sources inside the 
concealment with the normal materials, the wave field satisfies the Maxwell equation in the 3D whole space $R^3$ which includes the
anisotropic inhomogeneous cloak domain $\Omega _c$  and concealment  $\Omega _d$, and satisfies
the continuous interface conditions on the inner boundary surface $S_1$ and outer
boundary surface $S_2$  of the cloak domain $\Omega _c$.

Let $R_c  = R^3  - \Omega _c \bigcup {\Omega _d}$,  $R_d  = R^3  - \Omega _d$,
and by the EM integral equation (1), the EM wave field satisfies
\begin{equation}
\begin{array}{l}
 \left[ {\begin{array}{*{20}c}
   {E\left( {r,t} \right)}  \\
   {H\left( {r,t} \right)}  \\
\end{array}} \right] = \left[ {\begin{array}{*{20}c}
   {E_b \left( {r,t} \right)}  \\
   {H_b \left( {r,t} \right)}  \\
\end{array}} \right] + \\ 
 \int\limits_{\Omega _c \bigcup {\Omega _d } } {G_{E,H}^{J,M} 
\left( {r',r,t} \right) * _t \delta \left[ D \right]\left[ {\begin{array}{*{20}c}
   {E_b \left( {r',t} \right)}  \\
   {H_b \left( {r',t} \right)}  \\
\end{array}} \right]dr'}, \\ 
 \end{array}
\end{equation}
where $G_{E,H}^{J,M} \left( {r',r,t} \right)$ is the EM 
Green's tensor, its components
$E^J$, $H^J$, $E^M$, and $H^M \left( {r',r,t} \right)$ are the
EM Green's function on $\Omega _c \bigcup {\Omega _d } \bigcup {R_c} $, 
excited by the 
point  impulse sources outside of the concealment, $r \in R_d$.
By the assumptions, $G_{E,H}^{J,M} \left( {r',r,t} \right)$ exists on
$\Omega _c \bigcup {\Omega _d } \bigcup {R_c }$
and $G_{E,H}^{J,M} \left( {r',r,t} \right) = 0$, when $r' \in \Omega _d$.
The integral equation (5) becomes to
\begin{equation}
\begin{array}{l}
 \left[ {\begin{array}{*{20}c}
   {E\left( {r,t} \right)}  \\
   {H\left( {r,t} \right)}  \\
\end{array}} \right] = \left[ {\begin{array}{*{20}c}
   {E_b \left( {r,t} \right)}  \\
   {H_b \left( {r,t} \right)}  \\
\end{array}} \right] \\ 
  + \int\limits_{\Omega _c } {G_{E,H}^{J,M} \left( {r',r,t} \right) * _t \delta \left[ D \right]
\left[ {\begin{array}{*{20}c}
   {E_b \left( {r',t} \right)}  \\
   {H_b \left( {r',t} \right)}  \\
\end{array}} \right]dr'.}  \\ 
 \end{array}
\end{equation}

We consider the Maxwell equation in $R_d$, the
virtual source located $r$, $r \in R_d$ and the point source located $r_s$,  $ r_s  \in \Omega _d $ and $ r_s  \notin R_d$, we have
\begin{equation}
\begin{array}{l}
 \left[ {\begin{array}{*{20}c}
   {} & {\nabla  \times }  \\
   { - \nabla  \times } & {}  \\
\end{array}} \right]G_{E,H}^{J,M} \left( {r',r,t} \right) \\ 
  = \left[ D \right]G_{E,H}^{J,M} \left( {r',r,t} \right) + I\delta (r',r)\delta (t), \\ 
 \end{array}
\end{equation}
and
\begin{equation}
\begin{array}{l}
 \left[ {\begin{array}{*{20}c}
   {} & {\nabla  \times }  \\
   { - \nabla  \times } & {}  \\
\end{array}} \right]\left[ {\begin{array}{*{20}c}
   {E_b }  \\
   {H_b }  \\
\end{array}} \right]\left( {r',r_s,t} \right) \\ 
  = \left[ {D_b } \right]\left[ {\begin{array}{*{20}c}
   {E_b }  \\
   {H_b }  \\
\end{array}} \right]\left( {r',r_s,t} \right), \\ 
 \end{array}
\end{equation}
By using $ \left[ {E_b \left( {r,t} \right),H_b \left( {r,t} \right)} \right]$ to 
convolute (7), and $G_{E,H}^{J,M} \left( {r',r,t} \right)$ to convolute (8), 
to subtract the second result equation from the first result equation and make 
their integral in $\Omega _c \bigcup {R_c }$, and make integral by part and some manipulations, we can prove
\begin{equation}
\begin{array}{l}
 \left[ {\begin{array}{*{20}c}
   {E_b \left( {r,t} \right)}  \\
   {H_b \left( {r,t} \right)}  \\
\end{array}} \right] +  \\ 
  + \int\limits_{\Omega _c } {G_{E,H}^{J,M} \left( {r',r,t} \right) * _t \delta \left[ D \right]\left[ {\begin{array}{*{20}c}
   {E_b \left( {r',t} \right)}  \\
   {H_b \left( {r',t} \right)}  \\
\end{array}} \right]dr'}  \\ 
  =  - \oint\limits_{S_1 } {G_{E,H}^{J,M} \left( {r',r,t} \right) \times } \left[ {\begin{array}{*{20}c}
   {E_b \left( {r',t} \right)}  \\
   {H_b \left( {r',t} \right)}  \\
\end{array}} \right]d\vec S. \\ 
 \end{array}
\end{equation}

Because $G_{E,H}^{J,M} \left( {r',r,t} \right) = 0$, 
$ r' \in \Omega _d$, by continuous interface conditions of
$G_{E,H}^{J,M} \left( {r',r,t} \right)$, the term in right hand side of (9) is vanished, we have
\begin{equation}
\begin{array}{l}
 \left[ {\begin{array}{*{20}c}
   {E_b \left( {r,t} \right)}  \\
   {H_b \left( {r,t} \right)}  \\
\end{array}} \right] +  \\ 
  + \int\limits_{\Omega _c } {G_{E,H,b}^{J,M} \left( {r',r,t} \right) * _t \delta \left[ D \right]\left[ {\begin{array}{*{20}c}
   {E\left( {r',t} \right)}  \\
   {H\left( {r',t} \right)}  \\
\end{array}} \right]dr' = 0}. \\ 
 \end{array}
\end{equation}

Upon substituting integral equation (10) into the integral equation (6), we have
\begin{equation}
\left[ {\begin{array}{*{20}c}
   {E\left( {r,t} \right)}  \\
   {H\left( {r,t} \right)}  \\
\end{array}} \right] = 0.
\end{equation}

From the continuous property of the EM wave field, we obtain the following over vanish boundary
condition on the boundary $S_1$ of the concealment $\Omega _d$, we have
\begin{equation}
\left. {\left[ {\begin{array}{*{20}c}
   {E\left( {r,t} \right)}  \\
   {H\left( {r,t} \right)}  \\
\end{array}} \right]} \right|_{S_1 }  = 0.
\end{equation}

Because the EM wave field is excited by local sources inside of the concealment domain $\Omega _d$, it satisfies
the following Maxwell equation,
\begin{equation}
\begin{array}{l}
 \left[ {\begin{array}{*{20}c}
   {} & {\nabla  \times }  \\
   { - \nabla  \times } & {}  \\
\end{array}} \right]\left[ {\begin{array}{*{20}c}
   E  \\
   H  \\
\end{array}} \right]\left( {r',r_s,t} \right) \\ 
  = \left[ D \right]\left[ {\begin{array}{*{20}c}
   E  \\
   H  \\
\end{array}} \right]\left( {r',r_s,t} \right) + Q(r',r_s,t). \\ 
 \end{array}
\end{equation}

The equation (13) and its over vanish boundary condition (12) 
form contradiction equations on the concealment $\Omega _d$,
where 
$\left[ D \right] = diag\left[ {
   \varepsilon  \  \mu } \right](\partial /\partial t)$
with the normal EM material parameters $\varepsilon$ and $\mu$, $r_s  \in \Omega _d $ is
the source location, $Q(r',r_s,t)$ is the nozero local source inside of $ \Omega _d $.
There is no any Maxwell EM wave field to satisfy the contradiction equations (13) in
$\Omega _d$ with over vanish boundary
condition (12) on the boundary $S_1$. Therefore, we proved that ${ \bf there  \  is \  no\ Maxwell\  EM \  wave\  field\ excited \ by \  the}$
${\bf nonzero \ local\   sources\  inside\  of\  the\ cloaked\ }$ 
${\bf concealment.}$

In paper [1], we have used the simulations and theoretical analysis by the GL method to prove that 
the excited EM wave field by the point source outside of the sphere 
cloak [4]  through the cloak and never propagate enter the 
concealment and never be disturbed by the cloak. The assumptions and conditions in the theorem in this paper  are validated, 
therefore, we stress that there is no Maxwell  EM wave field excited by the nonzero local sources inside 
of the center sphere concealment cloaked by sphere cloak and the 
arbitrary geometry closed strip cloak. In practical cloak metamaterial fabrication 
and experiments, the phenomenon is presented in this paper should be received attentions, 
because the EM wave field excited in the concealment may be an
irregular EM chaos propagation in this region, it may interfere the EM devices and equipments 
working inside of the concealment, the high frequency irregular EM chaos radiation may
hurt the health of the human.

\begin{figure}[h]
\centerline{\includegraphics[width=0.86\linewidth,draft=false]{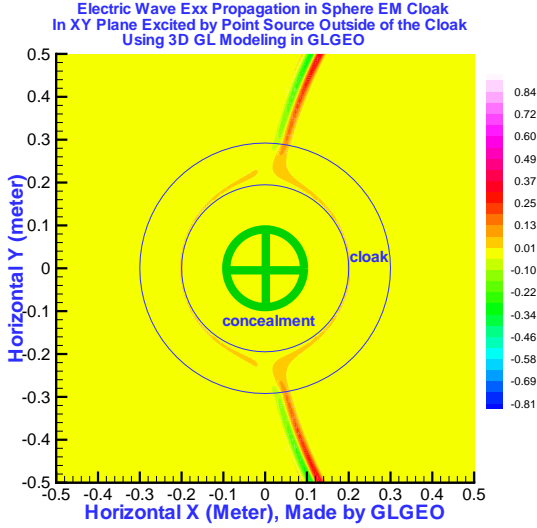}}
\caption{(color online) By the point source excitation in $x=1.1m$ on X axis outside of the cloak,
at the time step $110dt$,  the electric wave field $Exx$ propagation does 
disperse and split into the two phases around the concealment with an antenna $\oplus $, 
 and does not penetrate into it.}  \label{fig1}
\end{figure}

\begin{figure}[h]
\centerline{\includegraphics[width=0.86\linewidth,draft=false]{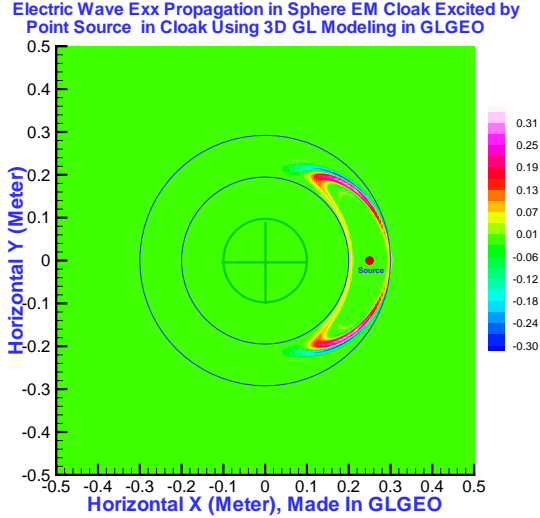}}
\caption{ (color online)  The electric wave field $Exx$ is excited by the point source inside of the cloak at x=0.25m 
on X axis, at the time step $15dt$, the $Exx$ propagates around the concealment and does not penetrate into it.}\label{fig2}
\end{figure}

\section{\label{sec:level1}Simulation of The EM Wave Field Through the Cloak By GL Method}
The simulation model:
the 3D domain is $ [-0.5m,0.5 m] \times [-0.5m,0.5 m]  \times [-0.5m,0.5m]$, 
the mesh number is $201 \times 201 \times 201$, the mesh size is 0.005m. 
The electric current point source is defined as
\begin{equation}
\delta (r - r_s )\delta (t)\vec e,
\end{equation}
where the $r_s$ denotes the location of the point source,
the unit vector $\vec e$ is the polarization direction, 
the time step $dt = 0.3333 \times 10^{ - 10}$ 
second, the  largest frequency $f=10 GHz$,  
the shortest wave length is $0.03m$.  
The EM cloak $\Omega _c $ is the spherical annular with the center in the origin and internal radius 
$R_1=0.2m$ and exterior radius $R_2=0.3m$. The cloak is divided into $50 \times 50 \times 50$ cells, 
the antenna subdomain $\oplus $, $\Omega _g$, is divided into 63 cells. The spherical coordinate is used in the sphere 
$r \le R_2$, the Cartesian rectangular coordinate is used in other where to mesh the domain. 

For a point source located outside of the cloak, the 3D GL EM modeling has been used to simulate the EM wave field through the sphere,
 ellipsoid, cylinder, and arbitrary closed strip complex geometry cloaks,
the simulations and theorems of the single and multiple sphere cloaks 
are proposed in paper [1].  The cloak simulations in papers [5-7] are 
proposed for the outside plane wave through cloak. The plane source to excite the 
plane wave can not be located inside of the cloak and concealment. Simulation for the EM wave field excited by the
point source, in particular, the source located inside of cloak, $r_s  \in \Omega _c$, or inside of 
concealment $r_s  \in \Omega _d$  is lack. The Figure 1 shows that the electric wave field $Exx$ is excited by
the current point source in the direction $\vec e = \vec x$ and located at $(1.1m,0.0,0.0)$ on the $X$ axis where is outside of the cloak. 
At the time step $110dt$,  the electric intensity wave field $Exx$ is propagating through the sphere annular cloak and around the 
sphere concealment, it does disperse and split into the two phases around the sphere concealment, 
the front phase speed exceeds the light speed; the back phase is slower than the light speed. 
The wave front outside of the cloak is the same as the exact $Exx$ propagation in free space, the 
$Exx$ wave field outside of the cloak never been disturbed by the cloak and never penetrate enter
the centre sphere concealment with the antenna $\oplus $.
The simulations of the EM wave field through the cloak excited by the point source inside 
of the cloak are presented in this paper. The electric wave field $Exx$ excited by the current point source in the direction 
$\vec e = \vec x$ and located at 
the point (0.25m,0,0) inside of the cloak is propagating at the time step $15dt$ that is shown in Figure 2. The Figure 3 
shows that the electric wave field $Exx$ propagates around the cloaked concealment at the time step $18dt$ , but does not penetrate into it
and its right wave front has been outside of the cloak. 
At the time step $48dt$, the electric wave field $Exx$ is propagating outside of the cloak that is shown in Figure 4,  however,
to compare an $Exx$  excited by the same source in free space, 
the electric wave field $Exx$ propagation outside of the cloak is disturbed by the cloak.
Figures 2 - 4 show that the wave field $Exx$  propagates through the cloak and never enter the concealment.
The concealment is complete concealed by the cloak from the EM wave field excited 
by the point sources inside of the cloak and in free space outside of the cloak. 
We did use the GL modeling to simulate many cases of the
EM wave field excited by the point sources inside of the concealment. However, all simulations are unstable and chaos. 
When the EM wave field is propagating to arrive the interface boundary $S_1$ between the concealment and cloak,
the GL simulation become unstable and chaos. The GL modeling simulation experiments 
reminder us to think may there is no any Maxwell EM wave field excited by the point source inside of the concealment. 
This is the motivation and idea of our theorem proposed in this paper. After the rigorous proof by the GL method, 
we obtained the theorem that  there is no Maxwell
EM wave field excited by the point sources located inside of the concealment.

\begin{figure}[h]
\centerline{\includegraphics[width=0.86\linewidth,draft=false]{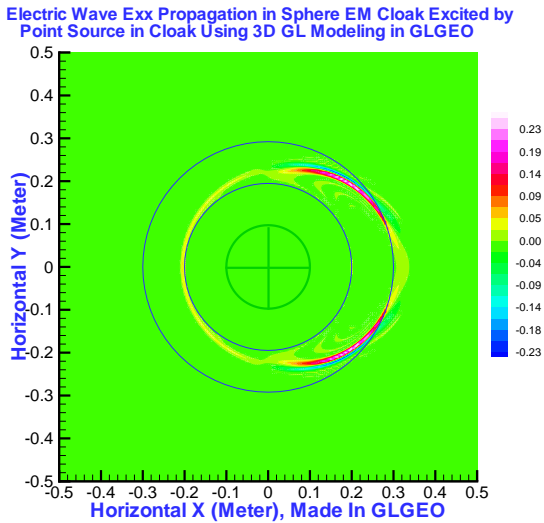}}
\caption{ (color online)  At the time step $18dt$, the $Exx$ propagates around the concealment and does not penetrate into it,
its right wave front is outside of the cloak}\label{fig3}
\end{figure}

\begin{figure}[h]
\centerline{\includegraphics[width=0.86\linewidth,draft=false]{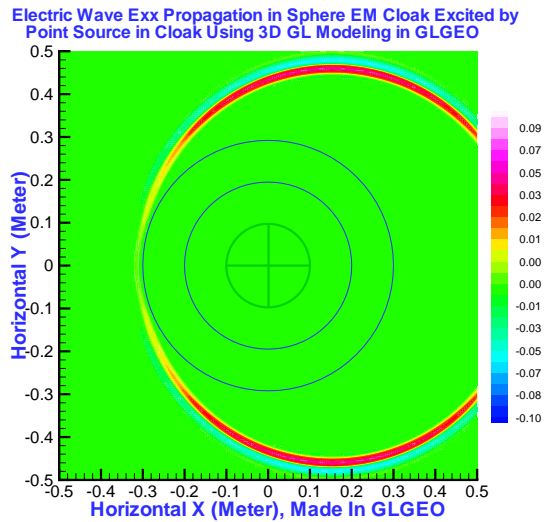}}
\caption{ (color online) At the time step $48dt$, the $Exx$ is propagating outside of the cloak,
the wave field propagation in the free space outside of the cloak is disturbed by the cloak.}\label{fig4}
\end{figure}

\section{\label{sec:level1}CONCLUSIONS}

The GL method is used to simulate the invisibility of the sphere and 
arbitrary cloaks and theoretically and rigorously proved theorem that there is no
Maxwell EM wave field excited by the nonzero local sources inside of 
the concealment which is cloaked by the sphere cloak and 
arbitrary closed strip cloak. 
A least square or regularizing chaos propagation of the EM wave field excited in the concealment
will be modeling and inversion by the GL metre carlo method [8] in next paper. However, any field
excited in the cloaked concealment can not be propagation outside of the concealment. 

The GL EM modeling is fully different from FEM and FD and Born Approximation
methods and overcome their difficulties. There is no big matrix equation to solve in GL method.
Moreover, it does not need artificial boundary and absorption condition
to truncate the infinite domain. The GL EM method consistent combines the
analytical and numerical approaches together. The GL method has double abilities of
the theoretical analysis and numerical simulations that
is shown in this paper.

The 3D GL simulations of the EM wave field through the single  and multiple sphere, cylinder, ellipsoid, and arbitrary geometry cloaks
show that the GLT and GLF EM modeling are accurate, stable and fast.
It saves more storages than the conventional methods and needs 10 to 50
minute to run 
the 3D EM wave field through the cloaks with 64 to 128 frequencies in the PC. The high performance
GL parallel algorithm in PC cluster
and super parallel computer is very fast and powerful to simulate
complex and large scale physical and chemical process.

The 3D and 2D GL parallel software is made and patented by GLGEO.
The GL modeling can be extended to its inversion \cite{8} and GL EM quantum field modeling
to solve quantization scattering problem of the electromagnetic field in the dispersive and loss metamaterials, cloaks
and more wide anisotropic materials.

\begin{acknowledgments}
We wish to acknowledge the support of the GL Geophysical Laboratory.
\end{acknowledgments}

\end{document}